\documentclass[preprint,superscriptaddress]{revtex4}

\newcommand{\ket}[1] {\left| #1 \right\rangle}

\begin{document}

\title{Reply to Comment on ``Properties and dynamics of generalized squeezed states''}

\author{Sahel Ashhab}
\affiliation{Advanced ICT Research Institute, National Institute of Information and Communications Technology (NICT), 4-2-1, Nukui-Kitamachi, Koganei, Tokyo 184-8795, Japan}
\affiliation{\small Research Institute for Science and Technology, Tokyo University of Science, 1-3 Kagurazaka, Shinjuku-ku, Tokyo 162-8601, Japan}

\author{Mohammad Ayyash}
\affiliation{Institute for Quantum Computing, University of Waterloo, 200 University Avenue West, Waterloo, Ontario N2L 3G1, Canada}
\affiliation{Red Blue Quantum Inc., 72 Ellis Crescent North, Waterloo, Ontario N2J 3N8, Canada}

\begin{abstract}
In our paper \cite{Ashhab2025}, our numerical simulations showed that, unlike displacement and conventional squeezing, higher-order squeezing exhibits oscillatory dynamics. Subsequently, Gordillo and Puebla pointed out that simulation results depend on whether the size of the state space in the simulations is even or odd \cite{Gordillo}. Using additional derivations, they argued that the oscillatory dynamics is unphysical and that the photon number must increase monotonically as a function of the squeezing parameter $r$. We agree with the observation of an even-odd parity dependence in the simulations. We independently noticed the same feature in our simulations after the publication of Ref.~\cite{Ashhab2025}. This observation led us to perform a more detailed investigation of the numerical simulation and mathematical aspects of the generalized squeezing problem. Our new findings were reported in Ref.~\cite{Ashhab2026}. Further analysis was reported in Ref.~\cite{Fischer}. Our conclusion is that the generalized squeezing operator is physically not well defined but can be made well defined when combined with additional information about the physical system under study. We demonstrated this point in the case where we include an additional nonlinear interaction term in the Hamiltonian. We disagree with the claim that the photon number must be a monotonically increasing function of $r$. This claim contradicts the mathematically rigorous results of Ref.~\cite{Fischer}. Furthermore, we show that the oscillatory behaviour persists in two closely related, well-behaved models.
\end{abstract}

\maketitle

\newpage

In our numerical simulations reported in Ref.~\cite{Ashhab2025}, although we performed a relatively large number of simulations with different simulation parameters, the size of the state space in the simulations was always an even number. Subsequently, the authors of the Comment \cite{Gordillo} and we \cite{Ashhab2026} noticed that the simulation results change drastically if we use odd simulation sizes. This parity dependence indicates that the squeezing operator and generally all physical quantities do not converge to well-defined asymptotic values in the limit of infinite simulation size. As a result, these quantities are physically not well defined.

In mathematical terms, the squeezing Hamiltonian is not essentially self-adjoint on the finite photon space \cite{Ashhab2026,Fischer}. The fact that various quantities converge to well-defined asymptotic values for the even-size or odd-size simulations indicates that these two cases (even and odd sizes) correspond to two different self-adjoint extensions of the Hamiltonian.

From a physical point of view, the fact that the simulation results do not converge in the limit of infinite simulation size indicates that the mathematical model is inadequate to describe a general physical situation. The statement that the high-order squeezing operator is not well defined can be interpreted as meaning that the operator does not contain sufficient information about the physical system under study to uniquely specify the dynamics. This situation can be intuitively understood as being somewhat similar to the situation where we are given a differential equation that describes the dynamical evolution of a propagating wave, but we are not given the boundary conditions. In this case, we will not know how the wave will be reflected when it reaches the boundary. The same equations of motion can lead to different dynamics, depending on the boundary conditions. One natural piece of information that can lead to a well-defined physical description of the generalized squeezing problem is the existence and details of higher-order nonlinear terms in the Hamiltonian. We showed in Ref.~\cite{Ashhab2026} how the simulation results become independent of simulation size when higher-order nonlinear Kerr interaction terms are included in the Hamiltonian.

The authors of the Comment \cite{Gordillo} performed a calculation of the average photon number using a power series expansion of an operator combination that contains the squeezing operator. Using this calculation, they concluded that the number of photons must be a monotonically increasing function of the squeezing parameter $r$. This calculation is interesting and might contain useful information. However, as was pointed out in Ref.~\cite{Fisher} and reiterated in Refs.~\cite{Ashhab2026,Fischer}, using standard power series expansions when dealing with the problem of generalized squeezing can lead to non-convergent series. The generating Hamiltonian for the generalized squeezing operator, $i \left[ \left(\hat{a}^{\dagger}\right)^n - \hat{a}^n\right]$, is not self-adjoint on the space spanned by the untruncated Fock state basis when $n\geq 3$. In other words, the Hamiltonian is physically not well defined on the physically intuitive state space spanned by photon number states ranging from zero to infinity. To make the squeezing operator unitary and physically meaningful, one must deal with subtle mathematical questions to find a domain (or state space) on which the Hamiltonian is self-adjoint. Similarly, one must be careful when performing power series expansions for this model, and one must in general provide convergence proofs to support the correctness of the conclusions. Any results obtained using non-convergent series expansions cannot be considered reliable. Our results in Ref.~\cite{Ashhab2026} show that the regularized dynamics that include a Kerr term are convergent and consistently exhibit oscillatory behaviour. Furthermore, our results in Ref.~\cite{Ashhab2026} strongly suggest that the spectrum of the squeezing Hamiltonian is discrete. The discreteness of the spectrum was proved rigorously by the authors of Ref.~\cite{Fischer}. According to the quantum recurrence theorem, if the spectrum is discrete, no observable can be monotonic indefinitely. Since the conclusion of the Comment \cite{Gordillo}, which is based on a heuristic series expansion, contradicts the results of rigorous mathematical analysis and the simulation-size-independent results of our recent simulations, we believe that the conclusion of the Comment is incorrect.

It is also interesting and relevant in this context to consider the quantum-pump model proposed in Ref.~\cite{Hillery}. In this model, we use the alternative squeezing operator:
\begin{equation}
\hat{U}_n \left( \tilde{r} \right) = \exp \left\{ -i \tilde{r} \hat{H}_{n, \rm QP} \right\},
\label{Eq:GeneralizedSqueezingOp_QuantizedPump}
\end{equation}
where
\begin{equation}
\hat{H}_{n, \rm QP} = i \left[ \hat{b} \left(\hat{a}^\dagger\right)^n - \hat{b}^{\dagger} \hat{a}^n \right],
\label{Eq:Hamiltonian_n_QuantizedPump}
\end{equation}
and $\hat{b}$ and $\hat{b}^{\dagger}$ are, respectively, the photon annihilation and creation operators of an additional electromagnetic field mode that acts as the pump field. In the following, we will express quantum states as $\ket{k_{\rm signal},k_{\rm pump}}$, where the first and second quantum numbers are, respectively, the signal and pump photon numbers. The reason why we introduced a different squeezing parameter ($\tilde{r}$) is that the effective squeezing parameter will now be a combination of $\tilde{r}$ and the initial number of photons in the pump field. In particular, let us assume for a moment that the pump field initially contains exactly $N$ photons. The Hamiltonian matrix element that connects the state $\ket{0,N}$ with the state $\ket{n,N-1}$ is then given by $\sqrt{n! \times N}$, i.e.~it acquires a factor $\sqrt{N}$ compared to the case where the operators $\hat{b}$ and $\hat{b}^{\dagger}$ are not present. The effective squeezing parameter is therefore given by $r=\tilde{r}\sqrt{N}$.

One important advantage of using the quantum-pump model is that it does not have any serious issues with self-adjointness, and hence we do not need to deal with subtle mathematical questions such as finding a self-adjoint extension defined on a carefully chosen domain \cite{Hillery}. We therefore eliminate a serious mathematical problem that plagues the generalized squeezing operator in the classical-pump model. Furthermore, this approach eliminates the need to choose a simulation truncation size arbitrarily. The initial number of photons in the quantum pump field introduces a natural cutoff: if the pump field runs out of photons, no more signal photons can be created. More specifically, the dynamics will be restricted to the state space $\left\{ \ket{0,N}, \ket{n,N-1}, \ket{2n,N-2}, \cdots, \ket{nN,0} \right\}$. Any simulation that contains all of these states will give the same results. Adding more quantum states with higher photon numbers in the simulation will not affect the results at all.

We can now ask what phenomena are obtained in this well-behaved model. These phenomena include the oscillatory dynamics, as well as the strong parity dependence. We have verified with numerical simulations that both of these phenomena appear in the quantum-pump model. The parity dependence is particularly remarkable in this context. No matter how large the initial number of photons in the pump field is, the ensuing dynamics depends strongly on whether this number is even or odd. It should be noted here that even in the case $n=3$, the average number of signal photons scales roughly as $N^{0.5}$ (since the scaling should be similar to that in Fig.~8(a) in our paper \cite{Ashhab2025}, even though the truncation there is done arbitrarily), such that the pump field is hardly depleted on average. Nevertheless, the dynamics is sensitive to the parity of the initial number.

If we make the commonly used assumption that the pump field is initially in a coherent state with a large average photon number, we find that this state contains almost equal probabilities of even and odd photon number states. As a result, the dynamics will involve a superposition of states with two drastically different oscillation patterns, one for the even numbers and one for the odd numbers. It is important to note here that it is hard to imagine any realistic experiment in which we start with a large number of photons in an electromagnetic mode and avoid the dissipative loss of a single photon in the finite duration of the experiment. Dissipative processes will therefore have a major effect on the dynamics. Furthermore, at a fundamental level, the model of a classical pump field as a coherent state is an approximation. For example, a laser that is used as a drive field in a quantum optics experiment is an open quantum system, with photons being continually added by the power source and lost to various loss mechanisms. In most physical situations, these fluctuations in the pump field have a small effect on the main phenomena being investigated. However, in the case of high-order squeezing, such incoherent processes would drastically affect the dynamics, because they mix the even and odd cases. These observations are all quite remarkable in their own right. We intend to investigate them and other aspects of the quantum-pump model in more detail in a future publication. The main point that we would like to make here is that the quantum-pump model, which avoids the subtle mathematical issues that plague the classical-pump model, also predicts oscillations in the dynamics. This result further confirms that the oscillations are not simply a mathematical artefact of flawed numerical simulations.

We would like to thank Daniel Braak, Daniel Burgarth, Felix Fischer and Davide Lonigro for useful discussions and for providing feedback for this reply. This work was supported by Japan's Ministry of Education, Culture, Sports, Science and Technology (MEXT) Quantum Leap Flagship Program Grant Number JPMXS0120319794.

\end{document}